# Optically induced currents in dielectrics and semiconductors as a nonlinear optical effect.

## Jacob B Khurgin[*]

*Johns Hopkins University, Baltimore MD 21218*
*\*Corresponding author: jakek@jhu.edu*

**We demonstrate theoretically that when well-below-the-bandgap femtosecond optical pulses propagate through a dielectric or semiconductor, DC current and charges are produced even though no real carriers are excited in the bands. The photo-induced current is a new ultra-fast nonlinear optical effect based on multi-photon quantum interference and creation of an asymmetric distribution of virtual carriers in the conduction and valence bands. We establish an unambiguous connection between the nonlinear optical conductivity responsible for the photo-induced DC currents and charges and the odd-order nonlinear optical *susceptibilities* of the material. We then apply our results to the recent experiments [1] in which photo-induced charges have been observed in $SiO_2$ irradiated by below-the-gap ultra-short optical pulses. Using a single well known measured value of the third order susceptibility (nonlinear index) of $SiO_2$ we obtain excellent agreement with all the experimental data of [1]. A clear physical picture of the origin of the photo-induced currents and charges shows that the versatility of ultrafast (virtual) nonlinear optical phenomena extends even further than had been previously thought.**

## 1. INTRODUCTION

In the more than half century since the first experimental [2] observations of optical frequency doubling and then mixing, nonlinear optics has evolved into an active field in which all kinds of frequency mixing processes over a wide spectral range, from UV to the THz and below have been not only demonstrated but put to practical use. The down-conversion of optical radiation into the low frequency domain is most interesting not only for its practical uses in generation of THz radiation [3] and in low noise optoelectronic oscillators [4], but also from the fundamental physics point of view, when the down-converted signal frequency $\omega_s$ approaches zero. For $\omega_s \neq 0$ one can equally well describe the down-conversion as a nonlinear polarization $P(\omega_s)$ generation or as a nonlinear current $J(\omega_s) = i\omega_s P(\omega_s)$ generation. However, when $\omega_s \to 0$ this relation obviously breaks down and one should be specific and careful in deciding whether current or polarization is generated first [5].

The lowest (second) order process in which the DC polarization is created is referred to as optical rectification (OR) [6]. It exists in materials without center of inversion symmetry (or biased with a DC field) and has been observed in many of them. Here we refer to the OR as a virtual (ultrafast) optical process in which the optical frequency is well below the absorption edge (bandgap) of the material and no real carriers are generated. Obviously, the slow equivalent of OR, involving generation of real carriers (electrons and holes) by the above-absorption-edge radiation is a process of photo-detection which is outside the scope of this article. It should only be noted here that while photo-detection can be either photoconductive (currents are generated) or photo-voltaic (polarization is generated), OR in all media, including semiconductors and their hetero structures [7-10] always leads to the appearance of a photo-induced voltage even though the term "virtual photoconductivity" has been used in [10], quite inappropriately in our view, to describe the appearance of nonlinear polarization in biased semiconductors.

When it comes to the unbiased materials having center of symmetry, the situation is different. A permanent polarization can be generated via third order optical susceptibility, when three photons have a well-determined phase relation. For instance, quantum interference of a path associated with absorption of two photons of a fundamental frequency $\omega$ with another path associated with absorption of a single photon of a second harmonic frequency $2\omega$ leads to DC polarization as $P_{DC} = \chi^{(3)}(0;\omega,\omega,-2\omega) : F_\omega F_\omega F_{2\omega}^*$ [12]. The observed effect can be real, when $2\hbar\omega$ is larger than the bandgap energy, or virtual in the opposite case. Both real and virtual effects were invoked in attempts to explain photo induced phase-matched second harmonic generation in fibers and nano-crystals in the 1980's and 1990's [13,14] Alternatively, it has been shown, that interference of $\omega$ and $2\omega$ waves can also lead to the appearance of directional photo-induced DC currents in fibers [15,16] single atoms [17], and semiconductors [18-21]. A coherent photo-injection tensor $\eta$ was introduced in [20] to describe the rate of increase in DC current as $\dot{J}_{DC} = \eta(0;\omega,\omega,-2\omega) : F_\omega F_\omega F_{2\omega}^*$. A very useful generalization was made in [22] where it was shown that directional current can be observed in any optical excitation in which average value of the cube of electric filed is not zero. *However, despite the obvious similarities between generation of directional currents and polarizations, no connection had been made at the time between these two effects, and furthermore, directional current photo injection was thought to be an exclusively real process, requiring $2\hbar\omega$ to exceed the bandgap [23].* That would have made coherent photo-injection a unique process, because for any other linear and nonlinear optical process with real excitation there exist its "virtual" counterpart related to it by Kramers-Kronig relation. For example, nonlinear absorption is always accompanied by the nonlinear refractive index. Hence in 1995 [24] the similarity between directional photocurrent and third order susceptibility was shown, and, as a consequence, it was predicted, phenomenologically, that even the non-resonant below the bandgap excitation of the semiconductor by a coherent superposition of fundamental and second harmonic waves should produce directional DC currents. These currents would be quite weak. Hence it is no wonder that they had not been observed and their existence had been disputed until 2013 when, as reported in [1], the DC currents had been generated in the fused silica excited by ultrashort optical pulses well below the bandgap.

The experimental arrangement used to observe the DC current and charges in [1] shown in Fig.1a is essentially identical to the one used in [22] to observe coherent photo-injection in GaAs, with a few femtosecond pulse exciting the $SiO_2$ sample between two contacts and

DC current measured in the external circuit. As in [22] a dispersive element had been used to control the relative phases of frequency components, and, depending on these phase relations, the polarity of the DC current had been shown to change. Despite the obvious similarity with the coherent photo-injection, the dependence of a photo-induced current observed in [1] did not fit the expected third-order curve and the explanation was made that involved Wannier Stark localization and essentially a complete change of the electronic structure of the $SiO_2$ adiabatically turning it into a conductor. The one dimensional model used in [1] was quite coarse and more refined calculations were performed in [25-27] that involved modeling the whole band structure of $SiO_2$ and then solving the density matrix equation. A connection with coherent photo-injection had been made but only the effect associated with photo-excitation of real carriers was invoked [1]. The most complete study [28] used an extensive band structure model and showed that the observed photocurrent in [2] has components associated with both virtual and real components, the latter being essentially the result of optical breakdown. None of the work [25-28] required the drastic change in electronic structure predicted in [1] and the fact that observations in [1] are related to coherent photo-injection was finally admitted in [29] but no simple theory relating the two has emerged up until now.

In this work we rigorously establish the presence of the virtual component of coherent photo-injection, predicted in [24] and thus related to the most basic of the optical nonlinearities – the nonlinear index of refraction. We show, that unlike the usual case of interference of two monochromatic waves, where the third order process always dominates, for short pulses, depending on the length, higher odd order processes become important and even dominant. By making a connection with the nonlinear index, we succeed in faithfully and quantitatively replicating all the observations made in [1] using just one well-known $SiO_2$ parameter – the value of the nonlinear index n2 without invoking any detailed model of band structure, without knowing the values of the strength of optical transitions, and without solving any differential equations.

## 2. OPTICAL BLOCH EQUATIONS

The framework in which we describe the charge generation is Optical Bloch equations (OBE's) [30] which were expanded to the semiconductors in works [31-35]. In these works, OBE's have been successfully applied to the cases of charge and spin currents in semiconductors and their heterostructures. Here we expand the treatment to the below the gap excitation by a single femtosecond pulse. Let us consider a pair of states in the conduction and valence bands characterized by the common wavevector k energies $E_c(\mathbf{k})$ and $E_v(\mathbf{k})$, and velocities $v_{c,v}(\mathbf{k}) = \hbar^{-1}\partial E_{c,v}(\mathbf{k})/\partial \mathbf{k}$ as shown in Fig.1b. In the presence of an optical field characterized by the vector potential $A(t)$ the Hamiltonian can be written in the momentum gauge as

$$\hat{H}(\mathbf{k}) = \hat{H}_0(\mathbf{k}) - \frac{e\hat{\mathbf{p}}\cdot \mathbf{A}}{m_0}$$

$$= \begin{pmatrix} E_v(\mathbf{k}) - e\mathbf{v}_v \cdot \mathbf{A}(t) & -e\dfrac{\mathbf{P}_{cv} \cdot \mathbf{A}(t)}{m} \\ -e\dfrac{\mathbf{P}_{cv} \cdot \mathbf{A}(t)}{m} & E_c(\mathbf{k}) - e\mathbf{v}_c \cdot \mathbf{A}(t) \end{pmatrix}, \quad (1)$$

Where $\hat{H}_0(\mathbf{k})$ is Hamiltonian in the absence of excitation, $\hat{\mathbf{p}}$ is the momentum operator, and $\mathbf{P}_{cv}(\mathbf{k}) = \langle c|\hat{\mathbf{p}}|v\rangle$ is the Kane matrix element of the interband transition between the Bloch states in the conduction and valence bands. It should be noted that using momentum gauge in which current rather than polarization is generated is appropriate for describing the experiments in [1]. Since the current in [1] flows in the closed circuit, the measured charge is an integral of that current which has nothing to do with polarization. Substituting (1) into the equation for the evolution of density matrix $\dot{\rho} = -i\hbar^{-1}[\hat{H},\rho]$ we obtain a set of differential equations for the matrix elements

$$\dot{\rho}_{cc} = -\dot{\rho}_{vv} = i\frac{e}{m_0}(\rho_{vc} - \rho_{cv})\mathbf{P}_{vc} \cdot \mathbf{A}(t) - \frac{\rho_{cc}}{T_1}$$

$$\dot{\rho}_{cv} = \dot{\rho}_{vc}^* = -i\omega_{cv}\rho_{cv} + i(\rho_{vv} - \rho_{cc})\frac{e}{m_0}\mathbf{P}_{vc} \cdot \mathbf{A}(t) \quad (2)$$

$$+ i\rho_{cv}e\mathbf{v}_{cv} \cdot \mathbf{A}(t) - \frac{\rho_{cv}}{T_2},$$

Where $\hbar\omega_{cv} = E_c(\mathbf{k}) - E_v(\mathbf{k}) \sim 9eV$ is the transition energy, $\mathbf{v}_{cv} = \mathbf{v}_c(\mathbf{k}) - \mathbf{v}_v(\mathbf{k})$ is the difference between velocities in the conduction and valence bands, and we have introduced phenomenologically the relaxation times $T_1$ and $T_2$. we introduce the usual optical Bloch vector (OBV) components, $x(t) = \text{Re}(\rho_{cv})$, $y(t) = \text{Im}(\rho_{cv})$, and $z(t) = \rho_{vv} - \rho_{cc}$, as well as two **time-dependent** Rabi frequencies – the interband one, $\Omega_{cv}(t) = (e/\hbar m_0)\mathbf{P}_{vc} \cdot \mathbf{A}(t)$, and the intraband one, $\delta\Omega(t) = (e/\hbar)\mathbf{v}_{cv} \cdot \mathbf{A}(t)$ which yields the OBE's [30],

$$\dot{x} = (\omega_{cv} - \delta\Omega)y - x/T_2$$

$$\dot{y} = -(\omega_{cv} - \delta\Omega)x + z\Omega_{cv} - y/T_2 \quad (3)$$

$$\dot{z} = -2\Omega_{cv}y + (1-z)/T_1$$

Comparing (3) with a usual set of OBE's for the localized states [30] that possess no velocity, shows that the additional term, $\delta\Omega(t)$ that indicates that the transition energy between the states gets modulated as electrons and holes accelerate and decelerate in the optical field. Obviously for the states with opposite wave-vectors $\mathbf{k}$ and $-\mathbf{k}$ shown in Fig.1. the signs of $\delta\Omega$ are opposite.

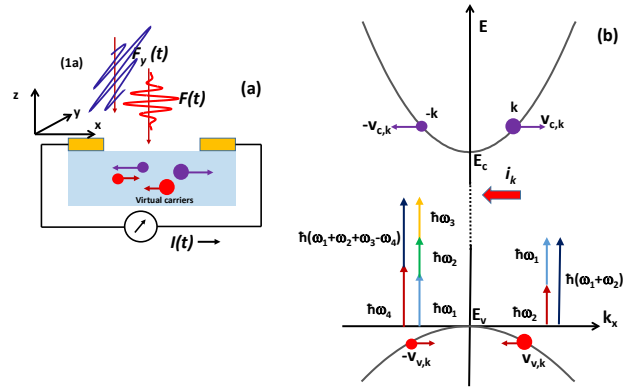

Fig.1. (a) Photo-injection experiment–geometry. The $SiO_2$ specimen has two unbiased contacts connected via the current meter. The incident optical pulse polarized along x (or a combination of x and y polarized pulses) excites current flowing in the external circuit. (b) Physical picture of the phenomena. Sub-bandgap few cycles optical pulse excites virtual electrons and holes in the bands, different for positive and negative values of k-vector, engendering non-zero electrical currents. Also shown are the diagrams explaining that non-zero currents are generated by quantum interference of one-photon and two-photon transitions (right), or two-photon and three-photon transitions (left)

## 3. ADIABATIC SOLUTION

Next we consider the adiabatic solution for short pulses, indicating that the pulse duration is much shorter than all relaxation times. This is obviously true when one compares the femtosecond duration of the pulses with the recombination time $T_1$, measured in hundreds of picoseconds. Furthermore, in semiconductors and dielectrics the below-the-gap absorption is exponential [36] rather than Lorentzian, meaning that the dephasing time $T_2$ is energy dependent and becomes very long away from the bandgap, making the material there effectively fully transparent (disregarding this fact may lead to erroneous result of having real carriers in the band caused by the light) Since the instant frequency of the optical pulse is much less than the optical transition frequency, $\dot{A}(t) \ll \omega_{cv} A(t)$, the dephasing time $T_2$ can also be safely assumed to be much longer than the pulse duration. The fact that that the instant frequency is small indicates that pulse passes through the medium adiabatically, and allows us to neglect the temporal derivative in the second equation of (3) to obtain

$$\begin{aligned}\dot{x} &= (\omega_{cv} - \delta\Omega)y \\ 0 &= -(\omega_{cv} - \delta\Omega)x + z\Omega_{cv} \\ \dot{z} &= -2\Omega_{cv} y\end{aligned} \quad (4)$$

With the initial condition $x(0) = y(0) = 0; z(0) = 1$ for relatively weak pulses we obtain the adiabatic solution

$$\begin{aligned}x(t) &= \Omega_{cv}(t)/(\omega_{cv} - \delta\Omega(t)) \\ y(t) &= \dot{\Omega}_{cv}(t)/(\omega_{cv} - \delta\Omega(t))^2 \\ z(t) &= 1 - \Omega_{cv}^2(t)/(\omega_{cv} - \delta\Omega(t))^2\end{aligned} \quad (5)$$

To confirm that (5) is indeed a solution of (3) we have solved (3) numerically for the femto-second optical pulse $F(t) = -F_0 \exp(-2\ln(2)t^2/T_p^2)\sin(\omega_0 t)$ polarized along x-axis and shown in Fig.2a with optical carrier $\omega_0 = 1.7 eV/\hbar = \omega_{cv}/5.7$ and with FWHM width $T_p = 3.6 fs$, or, in unit of optical cycles $N_{cyc} = T_p \omega_0 / 2\pi = 1.5$. The vector potential, found by integrating the electric field $A(t) = \int_{-\infty}^{t} F(\tau)d\tau$ is also shown in Fig.2a. in normalized units. This shape of course resembles $A(t) \approx (F_0/\omega_0)\exp(-2\ln(2)t^2/T_p^2)\cos(\omega_0 t)$. The relaxation times were taken to be $T_1 = 100 ps$ and $T_2 = 1 ps$ but it is easy to check that the adiabatic solution does not depend on these values. The maximum value of the Rabi frequency was taken to be $\Omega_{cv}(0) = \omega_{cv}/4$. Since the order of magnitude of $P_{cv}$ is $\hbar/a_b$, where $a_b = 1.6 A$ is the bond length, one can estimate $\Omega_{cv}(0) \approx eF_0/m_0\omega_0 a_b$ and hence the field corresponding to the above value of the Rabi frequency is on the scale of $F_0 \sim m_0 \omega_{cv} \omega_0 a_b / 4e \sim 10^{10} V/m$, i.e. a fraction of the intrinsic (or atomic) field that binds the crystal together, $F_a \sim \hbar\omega_{cv}/ea_b \sim 6\times 10^{10} V/m$, to which we shall return later on. Finally, the maximum value of the intraband Rabi frequency is taken to be $\delta\Omega = \pm\Omega_{cv}/4$, indicating that we consider the states $\pm k = \pm[k_x, 0, 0]$ (Fig.1) with velocities $v_{cv}(k) \sim \pm\hbar/4a_b m_0 \sim 2\times 10^5 m/s$. These states, according to fused silica band structure [37], are located roughly half-way between the Brillouin zone center and its edges. Thus all the parameters are realistic, and on the same scale as in experiments [1].

In Fig.2b we show the results for the OBV components for a positive value of $k_x$ and hence positive $\delta\Omega$. It is clear that OBV does follow the pulse adiabatically. For the state with opposite wave vector the results are only slightly different, and, not to crowd the plot we show that difference in the Fig.2c, where virtual electron-hole (e-h) pair populations for the states $\pm k$, $\rho_{cc,\pm k} = 1 - \rho_{vv,\pm k} = (1-z_k)/2$ are plotted. The expected value of the velocity for the given pair of states can then be found as

$$\begin{aligned}\langle v_{\pm k}\rangle &= Tr[\rho_{\pm k}\hat{v}] = \rho_{vv,\pm k}v_{v,\pm k} + \rho_{cc,\pm k}v_{c,\pm k} \\ &= \rho_{cc,\pm k}v_{v,\pm k} + v_{cv,\pm k} \\ &\approx v_{v,\pm k} + \tfrac{1}{2}\Omega_{cv}^2 v_{cv,\pm k}/(\omega_{cv} \mp \delta\Omega_{\pm k})^2\end{aligned} \quad (6)$$

One can see that the populations are not equal. Hence the net current flowing due to virtual e-h pairs in states $k$ and $-k$ is

$$i_k = -e(\langle v_k\rangle + \langle v_{-k}\rangle)/L = -e\Delta\rho_{cc,k}v_{cv,k}/L, \quad (7)$$

where $L$ is the distance between electrodes. The difference of the virtual e-h pairs $\Delta\rho_{cc,k} = \rho_{cc,k} - \rho_{cc,-k}$ is plotted in Fig. 2d. Expanding the denominator in (6) gives

$$\Delta\rho_{cc,k} \approx \frac{2\Omega_{cv,k}^2}{\omega_{cv,k}^2}\left[\frac{\delta\Omega_k}{\omega_{cv,k}} + \frac{\delta\Omega_k^3}{\omega_{cv,k}^3} + ...\right] \quad (8)$$

indicating that to the lowest order the virtual population difference and hence the current are expected to adiabatically follow the cube of the vector potential $A(t)$ as is indeed confirmed in Fig.2d where the normalized vector potential is also plotted. As the e-h pairs move between the electrodes the charge $Q_k(t) = \int_{-\infty}^{t} i_k(\tau)d\tau \sim \int_{-\infty}^{t} A^3(\tau)d\tau + ....$ flows in the external circuit as shown in Fig. 2e. Therefore, as long as the integrals of odd powers of vector potential are not zero a residual net charge $\Delta Q_k \sim \int_{-\infty}^{\infty} A^{2n+1}(\tau)d\tau$ will be measured at the end of each pulse, when, as evident from Fig.2c no carriers remain in the bands, as was indeed measured in the experiment [1]. Thus we have established that asymmetric (in momentum space) excitation of virtual e-h pairs generates real AC current that adiabatically follows a combination of odd powers of vector potential, and if the time integral of the current is not zero, a real charge is generated in the end of each pulse, and for the train of pulses a DC current flows in the external circuit.

It should be noted here that once the Rabi frequency peak value $\Omega_{cv}(0)$ approaches $\omega_{cv}$ the solution is no longer adiabatic as a finite population of real carriers remains at the end of the pulse – a tell-tale sign of optically-assisted tunneling described in [29], which, as its low frequency analog, the Zener effect, can be reversible. In this work we limit ourselves only to the instant (or virtual) effects occurring on the femtosecond scale.

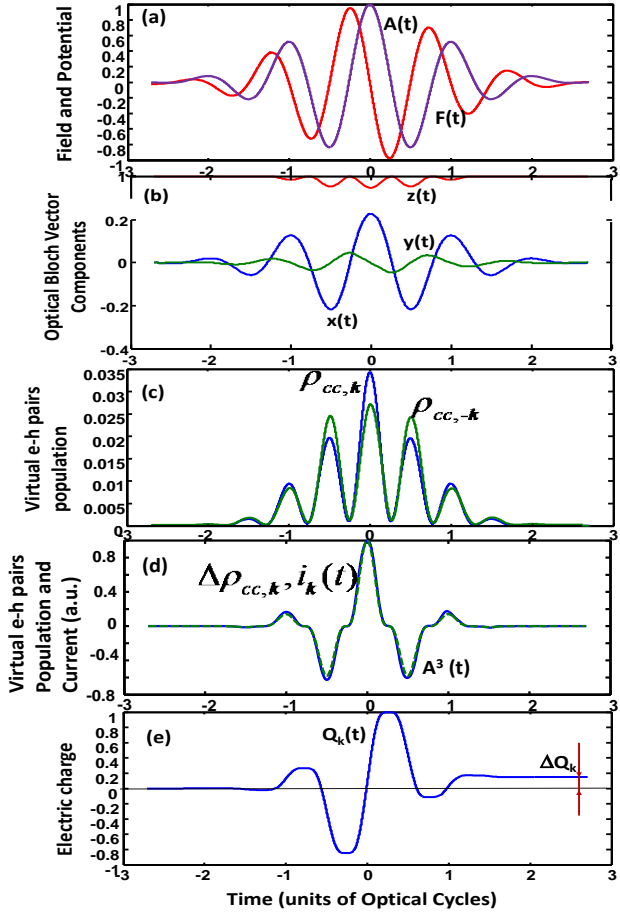

Fig.2. (a) Normalized electric field F(t) and vector potential A(t) of a few optical cycles pulse.(b) Adiabatic dynamics of the components of the OBV. (c) Dynamics of virtual e-h pair population of two states with opposite wave vectors **k** and –**k**. (d) Normalized difference between the above virtual hole populations, or, equivalently, normalized electrical current $i_k(t)$. Also shown as dashed line is the cube of the vector potential A³(t) (e) Electric charge passing through the external circuit showing non-zero remaining (accumulated) charge $\Delta Q_k$

## 4. NONLINEAR CONDUCTIVITY TENSOR AND ITS RELATION TO NONLINEAR OPTICAL SUSCEPBILITY

Having explained the origin of charge and DC current using just a pair of states with opposite wave-vectors we can now find the total current density by integrating over the whole Brillouin zone and performing summation over the bands

$$\boldsymbol{J}(t) = \frac{1}{4\pi^3} \sum_{c,v} \int \langle v_k \rangle d^3k$$

$$= \frac{1}{4\pi^3} \frac{e^4 A^3(t)}{m_0^2 \hbar^3} \sum_{c,v} \int \frac{\boldsymbol{v}_{cv,k} (\boldsymbol{v}_{cv,k} \cdot \hat{\boldsymbol{e}})(\boldsymbol{P}_{cv,k} \cdot \hat{\boldsymbol{e}})^2}{\omega_{cv,k}^3} d^3k +$$

$$+ \frac{1}{2\pi^3} \frac{e^6 A^5(t)}{m_0^2 \hbar^5} \sum_{c,v} \int \frac{\boldsymbol{v}_{cv,k} (\boldsymbol{v}_{cv,k} \cdot \hat{\boldsymbol{e}})^3 (\boldsymbol{P}_{cv,k} \cdot \hat{\boldsymbol{e}})^2}{\omega_{cv,k}^5} d^3k + ...$$

(9)

where $\hat{\boldsymbol{e}}$ is a unity vector indicating polarization of the optical wave. If we normalize vector potential to its maximum value

$\omega_0^{-1} F_0$ as $\boldsymbol{A}(t) = \omega_0^{-1} F_0 \boldsymbol{a}(t)$, we can introduce instant nonlinear optical conductivities as

$$\boldsymbol{J}(t) = \sigma^{(3)} : \boldsymbol{a}(t)\boldsymbol{a}(t)\boldsymbol{a}(t)F_0^3 \\ + \sigma^{(5)} : \boldsymbol{a}(t)\boldsymbol{a}(t)\boldsymbol{a}(t)\boldsymbol{a}(t)\boldsymbol{a}(t)F_0^5 + ...,$$

(10)

If we consider an isotropic material (the fact that fused silica has lower symmetry and thus additional nonlinear tensor components plays no role in further considerations) we obtain the following tensor components by performing averaging over the directions of the wavevector assuming an isotropic two-band model

$$\sigma_{xxxx}^{(3)} = \frac{1}{5\pi^2} \frac{e^4}{m_0^2 \hbar^3 \omega_0^3} \sum_{c,v} \int \frac{v_{cv,k}^2 P_{cv,k}^2}{\omega_{cv,k}^3} d^3k; \quad \sigma_{xxyy}^{(3)} = \frac{1}{3}\sigma_{xxxx}^{(3)}$$

$$\sigma_{xxxxxx}^{(5)} = \frac{1}{7\pi^2} \frac{e^4}{m_0^2 \hbar^5 \omega_0^5} \sum_{c,v} \int \frac{v_{cv,k}^4 P_{cv}^2}{\omega_{cv,k}^5} d^3k; \quad \sigma_{xxyyyy}^{(5)} = \frac{1}{5}\sigma_{xxxxxx}^{(5)}$$

(11)

The next logical step would be to assess the nonlinear conductivity by evaluating (11) which requires either a detailed knowledge of the full band structure (which is tedious), or using some crude model for it (which leads to significant errors). The summation over all bands is needed to remove the unphysical divergence at $\omega \to 0$, even though we are operating at frequencies above 350THz and this is not relevant.

Fortunately, the knowledge of band-structure is unnecessary if we can relate nonlinear conductivity to the experimentally-measurable characteristics of the material, in this case – nonlinear susceptibility [24]. It is well known that in all semiconductors and dielectric materials there exists a multi-photon absorption and associated with it nonlinear refractive index and susceptibility. One can write for the nonlinear polarization arising in the material,

$$\boldsymbol{P}(t) = \varepsilon_0 \chi^{(3)} : \boldsymbol{F}(t)\boldsymbol{F}(t)\boldsymbol{F}(t) \\ + \varepsilon_0 \chi^{(5)} : \boldsymbol{F}(t)\boldsymbol{F}(t)\boldsymbol{F}(t)\boldsymbol{F}(t)\boldsymbol{F}(t) + ...$$

(12)

where nonlinear susceptibilities can be estimated following [38-42] as

$$\chi_{xxxx}^{(3)}(\omega_0) = \frac{1}{5\pi^2} \frac{e^4}{\varepsilon_0 m_0^2 \hbar^3 \omega_0^4} \sum_{c,v} \int \frac{v_{cv,k}^2 P_{cv,k}^2}{\omega_{cv,k}^3} d^3k$$

$$\chi_{xxyy}^{(3)} = \frac{1}{3}\chi_{xxxx}^{(3)}$$

(13)

and similarly for the higher orders. We can thus make a connection between the nonlinear conductivity and susceptibility $\sigma^{(2n+1)} = \omega_0 \varepsilon_0 \chi^{(2n+1)}$.

The inherent connection between two processes is illustrated in Fig.3 using Feynman's diagrams. The diagram for a third-order process $\chi^{(3)}(\omega_2; \omega_1, \omega_2, -\omega_1)$ (cross-phase modulation) is shown in Fig.3a. In this process a virtual e-h pair is excited by the "pump" photon $\omega_1$; then these virtual carriers are moved in the field produced by the beating of pump and probe photons, and finally the pair recombines, producing polarization at the probe frequency $\omega_2$. For the coherent current injection $\sigma^{(3)}(0; \omega_1, \omega_2, -\omega_1 - \omega_2)$ (Fig.3b.) the three photon interference generates the DC current. The key point here is that all the band-to-band $P_{cv}/m_0$ and intraband $v_{c(v)}$ matrix elements of velocity involved in the processes (a) and (b) are identical, and for as long as all the energies are well below bandgap, the relation $\sigma^{(3)} = \omega_0 \varepsilon_0 \chi^{(3)}$ must be satisfied and the relations for higher order nonlinearities can be established in a similar manner.

Form this point on, there is no need to use Eq.(11) with all the difficulties associated with summation over large number of pairs of states as a clear connection with experimentally-measured material characteristics has been established.

If one introduces the intensity-dependent refractive index as $n(I) = n_0 + n_2 I + n_4 I^2 + ...$ where $n_2$ is the lowest order nonlinear index of refraction, it is easy to see that $\sigma^{(2n+1)} \sim n_{2n}$. Note, that here we have considered only odd order nonlinearities associated with multiphoton processes, and not the ones associated with Raman and AC Stark effects – this assumption is justified by the fact that the measured nonlinear index of fused silica (as well as in most semiconductors and dielectrics far from the bandgap) is positive. The third order susceptibility of fused silica is well known, and at wavelength of 805 nm in the order of $\chi^{(3)} \approx 2 \times 10^{-22} m^2/V^2$ [38] - a perfectly sensible number, since the order of magnitude of it is roughly $1/F_a^2$ where is the aforementioned atomic field. Note that the breakdown field is on the same scale as $F_a$ and is thus significantly higher than one used in [1] where the lattice constant instead of bond length was erroneously used to estimate the breakdown. This fact is important, as it indicates that the optical fields in [1] always remain lower by at least a factor of two than the fundamental breakdown field, and the observed breakdown is related only to the sample imperfections.

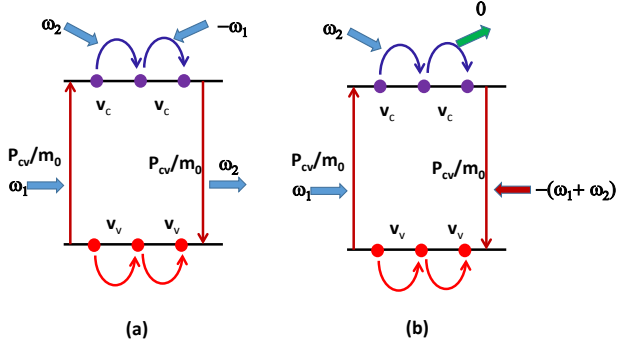

Fig.3. Feynman diagrams. (a) the third-order nonlinear optical process of cross-phase modulation $\chi^{(3)}(\omega_2; \omega_1, \omega_2, -\omega_1)$ in the two-band insulator or semiconductor (b) coherent photo-injection in the same material produced by interference of one and two photon transitions and described by a nonlinear conductivity $\sigma^{(3)}(0; \omega_1, \omega_2, -\omega_1 - \omega_2)$. The same matrix elements are involved in both cases.

## 5. PHOTOINDUCED CHARGE AND DC CURRENT

Assuming for simplicity that optical field is linearly polarized, if one now integrates (10), over the pulse length one obtains for the surface charge density that accumulates at the edges after each pulse

$$Q_{2D} = Q^{(3)} + Q^{(5)} + ...$$
$$= \sigma^{(3)} \omega_0^{-1} F_0^3 \langle a^3 \rangle + \sigma^{(5)} \omega_0^{-1} F_0^5 \langle a^5 \rangle + ... \quad (14)$$

where the normalized odd-power moment of vector potential is

$$\langle a^{2n+1} \rangle = \omega_0 \int a^{2n+1}(t) dt \quad (15)$$

Obviously, in the pulses that are physical the first order moment $\langle a \rangle = 0$. If ohmic contacts are attached to the edges and connected via an external circuit as in Fig.1a, the total charge that flows through that circuit is $Q = Q_{2D} A_{eff}$ where $A_{eff}$ the effective area where the current is generated. Since the higher order susceptibilities are related roughly as $\chi^{(3)}/\chi^{(5)} \approx \chi^{(1)}/\chi^{(3)}$ we finally obtain

$$Q = \varepsilon_0 \chi^{(3)} F_0^3 \left[ \langle a^3 \rangle + \left( \chi^{(3)} F_0^2 / \chi^{(1)} \right) \langle a^5 \rangle + \right.$$
$$\left. + \left( \chi^{(3)} F_0^2 / \chi^{(1)} \right)^2 \langle a^7 \rangle + ... \right] A_{eff} \quad (16)$$

This is the main result of our derivation and it is a remarkably uncomplicated result which contains just a couple of easily measured and thus well-known material parameters, $\chi^{(3)} \approx 2 \times 10^{-22} m^2/V^2$ [38], and, rather conveniently, $\chi^{(1)} = n^2 - 1 \approx 1$, so it can be disregarded. Unlike other theories this result requires neither knowledge of a full band structure, nor elaborate numerical calculations.

Now, the value of the atomic field obtained from $\chi^{(3)}$ measurement is $F_a \sim (\chi^{(3)})^{-1/2} \sim 7 \times 10^{-10} V/m$ which is understandably slightly different from the bond-length derived $F_a \sim 6 \times 10^{-10} V/m$. Furthermore, $\chi^{(3)}$ value reported in [43] is itself obtained by averaging the data from many experiments, showing at least 20-25% spread. Therefore, rather than choosing one or the other value we simply re-write (16) using the atomic field $F_a$ as a parameter that can be adjusted around these numbers to fit the experimental data the best. Getting ahead the best fit is obtained with $F_a \sim 5.36 \times 10^{-10} V/m$.

The re-written (16) now becomes

$$Q = \varepsilon_0 F_0 \left( F_0 / F_a \right)^2 \left[ \langle a^3 \rangle + \left( F_0 / F_a \right)^2 \langle a^5 \rangle + \right.$$
$$\left. + \left( F_0 / F_a \right)^4 \langle a^7 \rangle \right] A_{eff} \quad (17)$$

And it appears, at first glance, that for as long as $F_0/F_a < 1$ (which is to say all the way to the breakdown), the third order term shall dominate the higher order one, but that first impression is incorrect, because the odd-power moments of vector potential $\langle a^{2n+1} \rangle$ can vary by orders of magnitude depending on the FWHM of the pulse. For the Gaussian pulse $a(t) = \exp(-2\ln 2 \, t^2 / T^2) \cos(\omega_0 t)$

$$\langle a^{2n+1} \rangle \approx \frac{\pi \sqrt{(2n+1)\pi}}{2n\sqrt{2\ln 2}} e^{-\frac{\pi^2 N_{cyc}^2}{2(2n+1)\ln 2}} \quad (18)$$

Where $N_{cyc}$ is the number of optical cycles in the pulse's FWHM. The exact dependence is plotted in Fig.4a. Clearly, the magnitude of $\langle a^{2n+1} \rangle$ increases with (2n+1) and the relative magnitude of the higher order moments difference increases with the pulse length. For the pulses that are longer than 1.5 optical cycles the ratio of $\langle a^5 \rangle / \langle a^3 \rangle$ exceeds 10. Hence the fifth order contribution in (17) becomes dominant for $F_0 > F_a/3$ with the seventh order contribution taking over at $F_0 > 0.7 F_a$ and so on until breakdown.

This point can be easily understood if one applies Fourier transform to (10) – then, for as long as one operates well below the gap, the nonlinear conductivities are frequency-independent and one obtains (assuming for simplicity linearly polarized field) for the DC current (and charge)

$$J_{DC} \sim \sigma^{(3)} F_0^3 \int\int a(\omega_1) a(\omega_2) a^*(\omega_1+\omega_2) d\omega_1 d\omega_2 +$$
$$+ \sigma^{(5)} F_0^5 \int\int\int\int a(\omega_1) a(\omega_2) a(\omega_3) a^*(\omega_4)$$
$$a^*(\omega_1+\omega_2+\omega_3-\omega_4) d\omega_1 d\omega_2 d\omega_3 d\omega_4 + \ldots \quad (19)$$

The third order term is nothing but the virtual coherent photo injection current that originates from the quantum interference between the non-degenerate two photon absorption (2PA) of photons $\omega_1$ and $\omega_2$ and a one-photon absorption (1PA) of the sum frequency photon. The diagram of this process is shown in Fig. 1b, on the right. In all the works [18-23] only the degenerate effect was considered, i.e. interference between 2PA of fundamental photons $\omega$ and 1PA of the second harmonic $2\omega$, but the theoretical treatment had been extended to the non-degenerate case in [24]. Now, one can simultaneously find large numbers of photons of frequencies $\omega_1 \approx \omega_2$ and their sum frequency $\omega_1+\omega_2$ in the spectrum of optical pulse only when it roughly spans almost an octave, i.e. its length is no more than a couple of optical cycles – hence the sharp decay of the third order moment $\langle a^3 \rangle$ in Fig.4a.

The next, or fifth order term, originates from the interference of three-photon absorption of photons with frequencies $\omega_1 \approx \omega_2 \approx \omega_3$ and 2PA of the two photons whose frequencies are roughly in the range $\omega_4 \approx \omega_5 \approx 1.5\omega_1$ as also illustrated in Fig.1. on the left. Clearly, these 5 photons can be found in the spectrum of the pulse as long as it spans about one half-octave or more, hence the fifth order moment $\langle a^5 \rangle$ does not decay as fast as $\langle a^3 \rangle$ with higher order moments $\langle a^7 \rangle$ and $\langle a^9 \rangle$ continuing the trend.

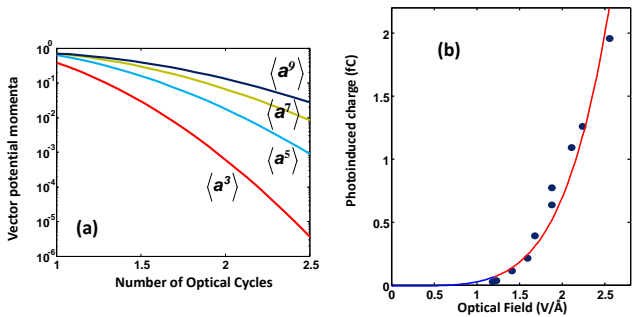

Fig.4. (a) Normalized averaged values of odd powers of vector potential as functions of the FWHM of optical pulse. (b) Photo-induced charge as a function of optical field obtained using Eq. (17) (solid line) compared with experimental results from ref [1] (solid circles) The change of color of the solid curve corresponds to the point where the fifth order term surpasses the third order term.

## 6. COMPARISION WITH EXPERIMENT

### A. Intensity dependence of the accumulated charge

We now apply our theory to the experiments performed in Ref. 1 with pulse width spanning 1.7 optical cycles. We evaluate the photo-induced charge (17) and fit the curve to the experimental data of Ref.1 treating $A_{eff}$ and $F_a$ as adjustable parameters. The best fit, shown in Fig 4b, is obtained for $F_a = 5.36 \times 10^{-10} V/m$ which is indeed close to the estimated value of the atomic field $(\chi^{(3)})^{-1/2} \approx 7 \times 10^{-10} m/V$, and for $A_{eff} = 2.3 \times 10^{-12} m^2$ which is quite reasonable for the geometry, and, as a matter of fact, differs from the one estimated in [1] by a factor of 2. The change of color of the curve roughly at $F_0 = 1.35 \ V/Å$ indicates the point at which the fifth order contribution surpasses the third order one and then remains the dominant effect up to the maximum amplitude, where the seventh order contribution becomes appreciable, but still much less than the fifth order one. One can also mention that the fit obtained using a simple analytical expression with a single experimentally parameter is superior to the ones obtained in [1] with detailed numerical modeling of band structure and a large number of assumptions being made. The fit is also comparable (with the exception of very high field where tunneling takes place) with the results obtained in full ab-initio calculations in [28].

### B. Propagation (dispersion) effects

As mentioned above, a few cycle optical pulse has broad, almost an octave-wide spectrum, hence, it's shape changes as it propagates through the medium whose dispersion can be characterized as $k(\omega) = k(\omega_0) + v_g^{-1}(\omega - \omega_0) + \frac{1}{2}\beta_2(\omega - \omega_0)^2$, where $v_g = \partial\omega/\partial k$ is a group velocity and $\beta_2 = \partial^2 k/\partial\omega$ is a second order (or group velocity) dispersion (GVD). The pulse envelope propagates with the group velocity while the phase of the optical carrier propagates with a different, phase velocity $v_p = k(\omega_0)/\omega_0$ - that leads to periodic changes of the profile of the shape of the $a(t)$ from symmetric to antisymmetric and back with a period characterized by the length $L_{2\pi} = \lambda|1-v_p/v_g|^{-1}$, as shown in Fig.5a, where we have plotted the shape of $a(t)$ and its third and fifth powers after the pulse propagated through the dispersive medium of lengths 0, $L_{2\pi}/4$ and $L_{2\pi}/2$ placed in front of the sample, as in Ref.1 As one can see the shape changes and the values of $\langle a^3 \rangle$ and $\langle a^5 \rangle$ oscillate accordingly. Obviously optically induced charge will be observed only when the shape of the vector potential is symmetric i.e. when the shape of the electric field is asymmetric. The symmetric electric field with $\langle F^{2n+1} \rangle \neq 0$ gives rise to DC polarization, while anti-symmetric electric field with $\langle a^{2n+1} \rangle \neq 0$ engenders DC current.

In addition to the periodic change of carrier phase caused by the velocities mismatch, GVD causes broadening of the envelope, characterized by the dispersion length $L_D = \beta_2^{-1} T_p^2$ at which the pulse length increases by a factor of two. In Fig.5b, using (17) with the previously determined values of $F_a$ and $A_{eff}$ and no fitting parameters we plot the value of the photo-induced charge as a function of propagation length, in units of $L_{2\pi}$ under the assumption of $L_D/L_{2\pi} = 3.3$ for the pulse with amplitude $F_0 = 1.7 V/Å$. Also shown are the experimental data from [1] – the agreement is obvious.

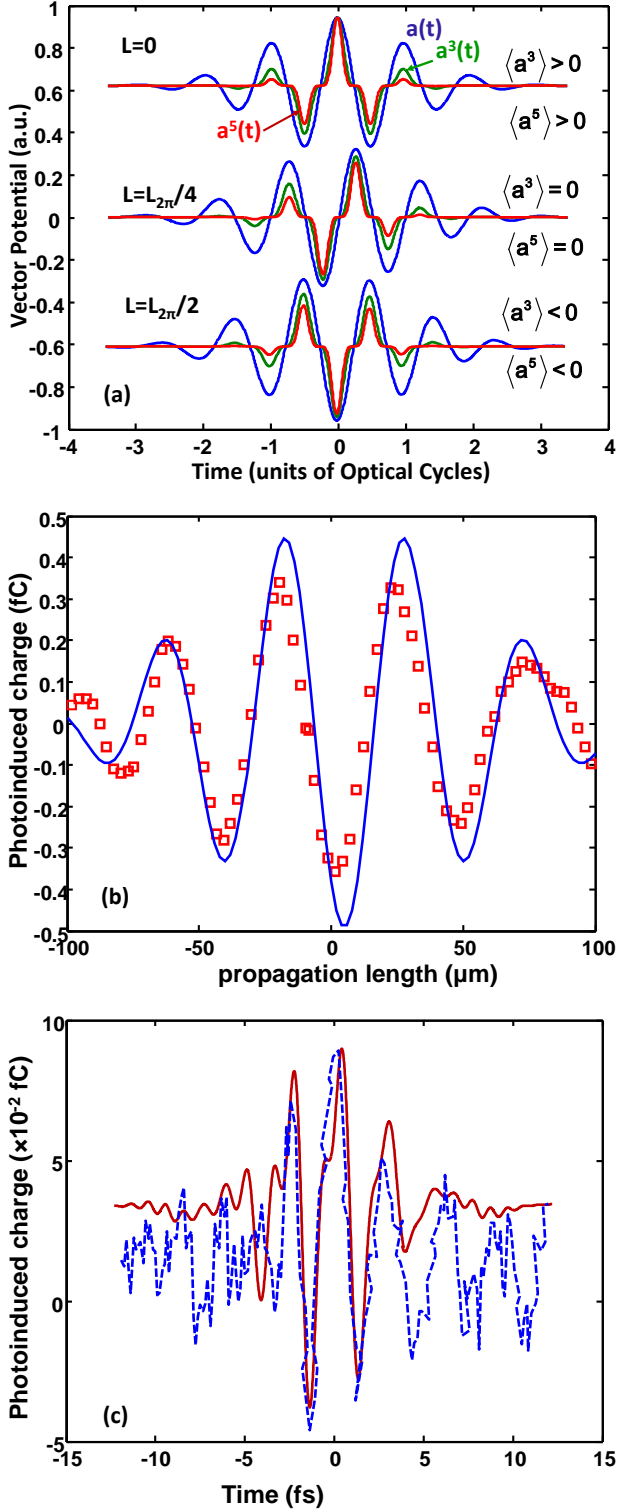

Figure 5. (a) Effect of ultra-short pulse propagation through the dispersive medium – the odd-power moments $\langle a^3 \rangle$ and $\langle a^5 \rangle$ exhibit oscillatory behavior as the phase shift between the carrier and the envelope changes. (b) Photo-induced charge as a function of the propagation distance through the dispersive medium leading to carrier-envelope phase shift. Squares are the experimental data from Ref.[1] (c) The photo-induced current produced by the orthogonally polarized injection (pump) and driving (probe) pulses as a function of pump-probe delay (solid line) compared with experimental data from Ref.[1] (dashed line).

### C. Cross-polarized pump-probe photo-injection

Among the most interesting experiments performed in [1] was the one in which two orthogonally polarized pulses – the strong pump (or "injection") pulse with amplitude $F_{y0}$ polarized along y direction and the weaker probe ( "driving" ) pulse with $F_0$ polarized along the *x* direction and delayed (Fig.1a) were combined on the sample. Both pulses passed through a SiO$_2$ plate of thickness $L_{2\pi}/4$; hence no photo-induced charge was generated unless the two pulses where delayed by amount $\Delta t$. While one way to describe this experiment is in terms of virtual carrier injection followed by drift in the field of the driving pulse, a much simpler explication can be obtained by invoking non-diagonal elements of nonlinear conductivity tensor elements $\sigma^{(3)}_{xxyy}, \sigma^{(5)}_{xxyyyy}$, and so on, whose relation to diagonal elements had been established in Eq. (11). Following each step in the derivation of (17), the photo-induced charge for the cross-polarized case can be found as

$$Q(\Delta t) = \varepsilon_0 F_0 \left(F_{0y}/F_a\right)^2 \left[\tfrac{1}{3}\langle a^{1+2}(\Delta t)\rangle + \tfrac{1}{5}\left(F_{0y}/F_a\right)^2 \langle a^{1+4}(\Delta t)\rangle + \tfrac{1}{7}\left(F_{0y}/F_a\right)^4 \langle a^{1+6}(\Delta t)\rangle\right] A_{\mathit{eff}} \quad (20)$$

Where the "cross-correlation moments" are

$$\langle a^{1+2n}\rangle(\Delta t) = \omega_0 \int a(t) a^{2n}(t - \Delta t) dt \quad (21)$$

The results obtained using Eq. (20) are shown in Fig.5c, where the experimental data is also shown. A good agreement indicates that photo-induced charge is indeed a nonlinear optical phenomenon.

### 7. CONCLUSIONS

In this work we have rigorously shown that a well-known effect of coherent photo-injection [16-24] in which DC currents are generated due to odd-order multiphoton quantum interference can be non-resonant. In other words, even when two or three photons do not have energy sufficient to produce real carriers, virtual carriers are being generated and their asymmetric distribution in k-space produces real currents flowing in the external circuit. Provided proper phase relations within the Fourier spectrum of optical excitation (or, equivalently, between carrier and envelope of the pulse), a non-zero DC current flows through the external circuit. While the prediction of this "virtual" coherent photo-injection was made in 1995 [24], it remained untested until the experiments in [1] which this theory explains very well, both qualitatively and quantitatively.

What is more, we have firmly established that coherent photo injection is nothing but an odd order nonlinear optical effect, closely related to the nonlinear refractive index. This effect relies on delocalized states in the conduction and valence bands; hence it must be described in the momentum gauge using nonlinear optical conductivity, rather than susceptibility, but this is just a nonlinear optical effect.

A peculiar point of the coherent current photo-injection established by us is that, depending on the bandwidth of the optical field, different terms in perturbative expansion (17) tend to dominate, so the power dependence of photo-induced charge cannot be described by a single $F_0^{2n+1}$ term over the sufficiently wide range of optical powers. But the effect is still perturbative in its nature, just higher order terms do play a significant role.

There are a couple of interesting issues that require additional attention of the community. At first glance, it is not clear where does the energy required to move the charge through the resistive external circuit come from? In our view, the current itself modifies the effective refractive index which causes the adiabatic red shift of the pulse. Detecting this shift would be an interesting tasks for experimentalists. Another issue is concerned with exactly what happens at the contacts? In our view the electron in the valence band can be adiabatically transferred into the metal electrode via the conduction band without staying in that band. More detailed theoretical study of this effect is needed.

What is perhaps most relevant to the experimental community is that we have obtained a strikingly simple relation between this virtual coherent photo-injection and the well-known nonlinear index of refraction whose value can be reliably measured. This value (in addition to the well-known value of linear refractive index) turned out to be sufficient to quantitatively explain all the results in [1] without making any assumptions about the band-structure and the strength of optical transitions. Therefore, knowing the nonlinear index, one can obtain a realistic estimate of the photo-induced charges and currents in any material without resorting to tedious calculations. In retrospect, this result is not surprising, as all the off-resonant nonlinear processes in solids can be perfectly well described using the most simplistic bond orbital models [44]. Yet this result serves as a reminder that the range of nonlinear optics keeps expanding and many seemingly exotic phenomena can be explained using nonlinear optics formalism

The author would like to thank Prof. Vladislav Yakovlev for stimulating e-mail correspondence and Dr. P. Noir for his thought-provoking questions asked.